\documentclass{ab}
\usepackage{graphicx}
\usepackage{natbib}

\begin{document}  

\title{Study of Distortions in Statistics of Counts 
in CCD Observations using the Fano Factor}

\author{{I.~V.}~{Afanasieva}}

\institute{Special Astrophysical Observatory, Russian Academy of Sciences, Nizhnij Arkhyz, 369167 Russia}

\institute{ITMO University, St. Petersburg, 197101 Russia}

\titlerunning{Study of Distortions in Statistics of Counts in CCD Observations using the Fano Factor}
\authorrunning{Afanasieva}

\date{May 24, 2016/Revised June 17, 2016}
\offprints{Irina Afanasieva \email{riv@sao.ru}} 

\abstract{
Factors distorting the statistics of photocounts when detecting
objects with low fluxes were considered here. Measurements of the
Fano factor for existing CCD systems were conducted. The study allows
one to conclude on the quality of the CCD video signal processing
channel. The optimal strategy for faint object observations was
suggested.
}

\maketitle
 
\section{INTRODUCTION}

There is an opinion formed among astronomers that charge-coupled
array devices are almost ideal light detectors which do not distort
the input Poisson signal, and observations with them are restricted
only by the statistics of the original flux.

The distribution of counts during image acquisition on the whole
differs from the distribution of photons that causes it. It is
conditioned by many factors:

\begin{list}{}{
\setlength\leftmargin{2mm} \setlength\topsep{2mm}
\setlength\parsep{0mm} \setlength\itemsep{2mm} } \item readout
noises; \item inhomogeneity of the array sensitivity; \item
instability and non-linearity of the transfer function; \item
cosmic ray traces and interference effects on thinned sensors
(fringes).
\end{list}

If it refers to faint object observations on a relatively bright sky
background, then the list of factors can be completed with correct
background subtraction which also distorts the statistics of
photocounts and, consequently, lowers the signal-to-noise ({\it S/N})
ratio of the final result.

The aim of the present paper is to reveal restrictions imposed by the
above factors on low-flux objects acquisition with CCDs.

\section{DISTORTIONS OF CCD COUNTS' STATISTICS}

As a rule, when operating with CCD a hypothesis is accepted that the
statistics of output counts follows the Poisson
law~\citep{Howell2006:Afanasieva_n}. This very reason is used in
determination of the gain factor ({$gain$}) of an analog-to-digital
converter in the CCD readout channel. Usually a sequence of image
pairs of uniformly illuminated field with different exposures {$t$}
is used. In this case, the measured dispersion of sequence of counts
in the image difference $\Delta I(x_i, y_i, t)$ is determined by the
relation:

\begin{equation}
D_I(t) = {( \overline{\Delta I(t)^2} - \overline{\Delta I(t)}^2 ) / 2}.
\end{equation}

In the case of the Poisson law, the dispersion of counts will be
proportional to the mean value:

\begin{equation}
D_I(t) = gain \times \overline{I(t)}.
\end{equation}

Naturally, realistic dependences of dispersion on the mean differ
from the linear ones. At small counts---as a consequence of readout
noise, at greater---due to non-linearity of the transfer function of
the CCD readout channel. It is known that one of the strongest
testing criteria of the Poisson distribution of counts is connected
with the study of the so-called Fano
factor~\citep{fano47:Afanasieva_n} (dispersion index, variation
factor~\citep{cox66:Afanasieva_n}) $k(t)$ which is the relation of
dispersion to the mean value:

\begin{equation}
k(t) = {D_I(t) / \overline{I(t)}}.
\end{equation}

\begin{figure*}[tbp!!!]
\centerline{\includegraphics[width=160mm]{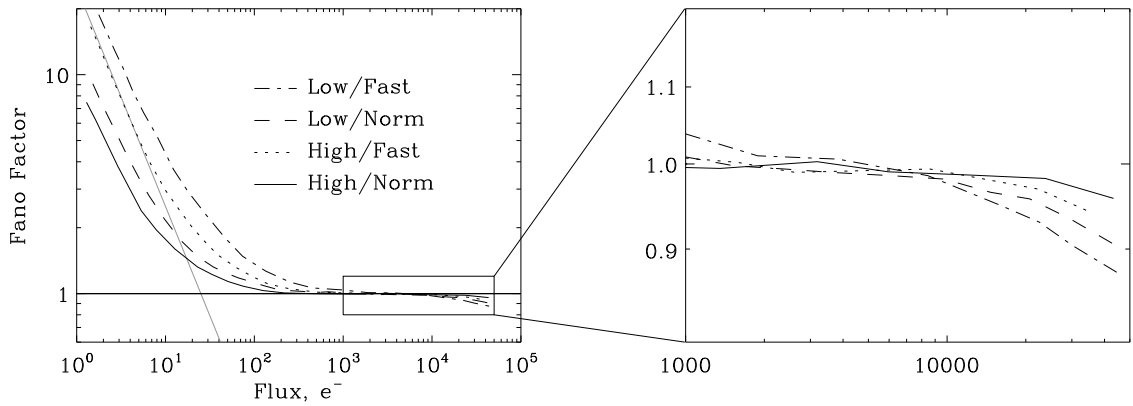}}
\caption{Dependence of the Fano factor in the flat field image on the
average flux for various operation modes of the CCD42-90 readout
channel: {\it Low} and {\it High} are {$gain$} values equal to
$1.82e^-$/ADU and $0.48e^-$/ADU respectively; {\it Fast} and  {\it
Norm} are readout rates, 400 and 100 Kpixel s$^{-1}$ respectively. An
oblique grey straight line shows the asymptotic behavior in the
region of low fluxes for the {\it High/Fast} mode with readout noise
(RON)~$5e^-$. } 
\label{fig1:Afanasieva_n}
\end{figure*}

For the Poisson distribution, $k(t)\equiv1$. To check the difference
between the CCD and Poisson distributions of counts, we measured the
Fano factor for existing dCCD
systems~\citep{mar00:Afanasieva_n,afa15:Afanasieva_n} designed by the
Advanced Design Laboratory and operated at the 6-m SAO RAS
telescope at present.
Figure~1 shows the measurements for the CCD
readout channel with EEV CCD42-90 detector.

As can be seen from Fig.~1, the differences of
the Poisson distribution are most noticeable at low fluxes and are
determined by readout noise and bandwidth of a video channel:

\begin{equation}
k(t) \approx { RON^2 / \overline{I(t)}}.
\end{equation}

Inconsiderable fall of the Fano factor in the greater-flux region
appears due to the dispersion decrease which is discussed in recent
works~\citep{Down06:Afanasieva_n,Ma2014:Afanasieva_n,Guyonnet15:Afanasieva_n}
and is probably associated with a charge redistribution between
pixels. This effect is one more factor limiting accuracy at high
levels of signal in pixel.

Thus, one can conclude the fact that the count statistics in certain
pixels of a realistic CCD is close to the Poisson one in a relatively
wide range of signal intensity.

However, in spite of the fact that the CCD detector system does not
practically distort the Poisson statistics in each pixel, it is
noteworthy that still there are distortions in the acquired image at
different spatial frequencies. First, sensitivity variations (quantum
efficiency nonuniformity) of certain pixels can cause
this~\citep{ter81:Afanasieva_n}. Second, the distortions are caused by
interference effects which occur on a thin layer of a substrate in a
back-illuminated CCD~\citep{les90:Afanasieva_n}, so-called fringes.
Their amplitude depends on wavelength and substrate thickness. To
estimate the count statistics distortions, we obtained an image of a
flat field with the EEV CCD42-90 detector in the spectral range of
760--920~nm. In this case, a root-mean-square amplitude of image
modulation by fringes was approximately 2.6\% and the modulation
period was about 30~pixels. Then a sample of random series of centers
$\{x_i,y_i\}$ of square fragments ({\it box}) with different sizes
{$w$} and a volume of about $10^4$ was generated. The Fano factor
$k_i (w)$ was calculated in each fragment, and the selective Fano
factor $\overline{k(w)}$---over the whole sequence.
Figure~2 shows the dependence of the selective
Fano factor on the fragment size, which demonstrates the count
statistics distortion at different spatial frequencies. The same
figure demonstrates the values for a fringe-free image in accordance
with~\citep{how12:Afanasieva_n}.

The fact that the Fano factor slightly increases in the fringe-free
image gives evidence of small distortions of the count statistics.

\section{COUNT STATISTICS DURING FAINT OBJECT DETECTION}

\begin{figure}[tbp!!!]
\includegraphics[width=80mm]{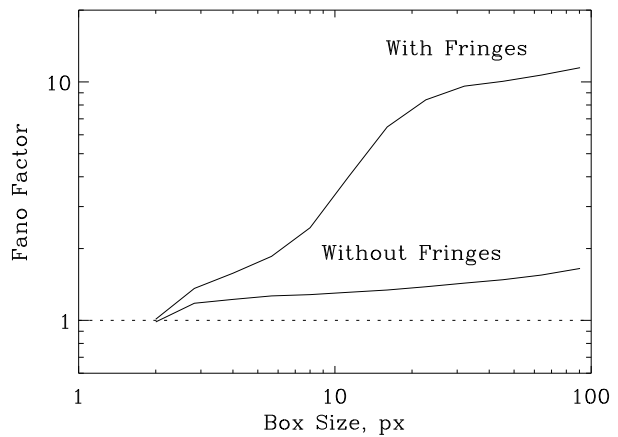}
\caption{Dependence of the selective Fano factor on the fragment size
before and after the removal of fringes for~CCD42-90. A dashed line
indicates the value characteristic for a Poisson flux.}
\label{fig2:Afanasieva_n}
\end{figure}

Let us assume that the incoherent radiation flux \mbox{$F=F_{\rm
obj}+F_{\rm sky}$}, which is the sum of fluxes from a studied object
and the sky background and \mbox{$F_{\rm obj} \le F_{\rm sky}$}, is
falling on the input of the registration system by which we mean the
system {\it atmosphere~+~telescope~+~spectrograph~+~CCD}. With
exposures exceeding the time of atmospheric coherence, the influence
of the latter can be neglected; and it is a quite true hypothesis
that a flux at the registration system entrance is a random process
with a Poisson distribution which becomes close to a Gaussian
distribution~\citep{mon07:Afanasieva_n} in the case of great fluxes. Then, the
measured signal at the output of the registration system at the point
with coordinates $(x,y)$ can be set with the expression:
\begin{equation}
N(x,y) = F(x,y)  flat(x,y) + bias,
\end{equation}
where $bias$ is a zero level of a CCD system and $flat(x,y)$ is a
flat field function determining the variation of a CCD transfer
function over the field of view. The $bias$ level is a random value
and its root-mean-square error is readout noise of a CCD system.

Hereinafter, by faint objects we mean objects both with low flux and
detectable against the sky background with some contrast $c$:

\begin{equation}
c = { ( F_{{\rm obj}+{\rm sky}} - F_{\rm sky} ) / F_{\rm sky}}.
\end{equation}

For example, the object fainter than the 20th magnitude in the
visible range with typical images at the 6-m telescope
(1\farcs5--2\arcsec) should have a contrast of $1$.

During actual observations of faint objects, two independent values
are measured: $N_{{\rm obj}+{\rm sky}}$ and $N_{\rm sky}$, which are
defined with the relations:
\begin{equation}
\begin{array}{lcl}
N_{{\rm obj}+{\rm sky}} = (F_{\rm obj} + F_{\rm sky})  flat(x,y) + bias_{{\rm obj}+{\rm sky}},\\
N_{\rm sky} = F_{\rm sky}  flat(x,y) + bias_{\rm sky},\\
N_{\rm obj} = N_{{\rm obj}+{\rm sky}} - N_{\rm sky}.
\end{array}
\end{equation}

As a matter of fact, these expressions connect an observed
mathematical expectation of a number of counts of the studied object
with mean measured values. Let us note that the     transfer function
$flat(x,y)$ in each observing run is determined with some error,
thus, it is also a random value. Our laboratory measurements show
that the realistic distribution of estimation errors of the flat
field and readout noise of a readout channel are well presented with
the Gaussian distribution.

The ultimate aim of observations is not only measuring mean values
but also the reconstruction of signal statistical properties from the
analysis of count statistics at the output of a readout channel which
introduces distortions. Solving of this problem in general terms
seems quite difficult and is not a goal of the present paper.

To estimate the statistics distortions in an existing CCD system, we
carried out a numerical modeling with the Monte-Carlo technique. It
was supposed that a random photon flux with a Poisson distribution
fall on the CCD system input. It was also accepted that estimation
errors of the flat field and readout noise have a Gaussian
distribution. We studied the case of observation of the object with a
contrast of \mbox{$c=1$} \mbox{$(F_{\rm obj}=F_{\rm sky})$}. The
object's  Fano factor for a random number sample of a volume of
$10^4$ instances was calculated at each sky background level.

Figure~3 shows the dependence of the Fano
factor on the sky background level for different errors in the
flat-field determination. Readout noise of $3e^-$ was accepted in
calculations. As the figure indicates, the Fano factor is minimal at
a certain level of the sky background, i.e., in this case, the CCD
system introduces minimum distortions of the counts statistics and,
correspondingly, a signal-to-noise ratio in this region is maximum.
The increasing of the Fano factor on low fluxes is obviously
determined by the readout noise of a CCD readout channel.

\begin{figure}[tbp!!!]
\includegraphics[width=80mm]{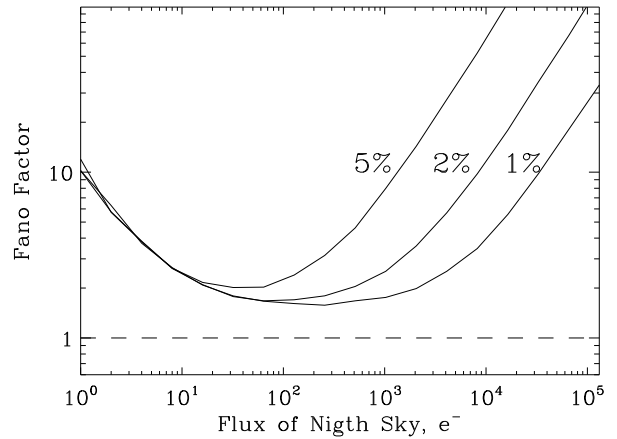}
\caption{Dependence of a Fano factor on a level of the sky background
for different errors in the flat-field determination---1\%, 2\%, and
5\%.  A dashed line indicates the value characteristic for a Poisson
flux.} \label{fig3:Afanasieva_n}
\end{figure}

Increasing of the Fano factor with the growth of the sky background
brightness means that at a high signal level, the noise in an image
is determined by uncertainty of a flat field value. Consequently, a
signal-to-noise ratio of a measured object does not grow with the
increase of a sky background flux. It is shown in
Fig.~4, which demonstrates the result obtained
with a numerical modeling using another units. This result is
important to get the maximum {\it S/N} ratio during observations of
faint objects using actual CCD detector systems with a readout noise,
and in an inhomogeneous sensitivity correction procedure is used. It
is noteworthy that an error in the flat field determination depends
not only on a signal level of calibration images but also on the
presence of residual noise after removal of fringes, unavoidable dust
particles on the optical elements of a registration system (an focal
reducer or a spectrograph), and on the violation of telecentric
conditions in a calibration path, etc. The influence of the last
factor can be essentially reduced by using images or spectra of
twilight sky for calibration.

\begin{figure}[tbp!!!]
\includegraphics[width=80mm]{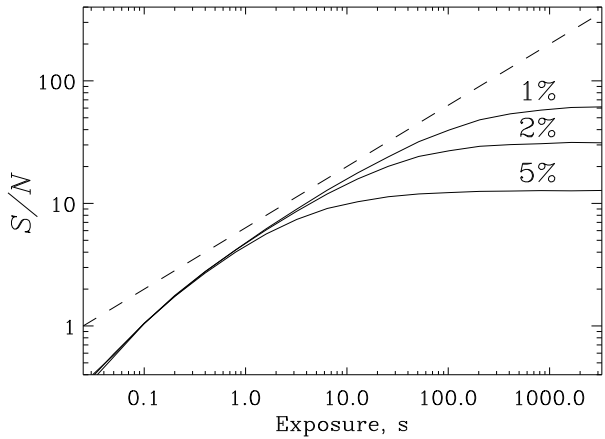}
\caption{Signal-to-noise ratio for an object depending on exposure
for different errors in the flat-field determination---1, 2, and 5\%.
A dashed line indicates the {\it S/N} value for a Poisson flux.}
\label{fig4:Afanasieva_n}
\end{figure}

Dependences shown in Fig.~3 allow us to set the
limiting values of optimal exposures for observations with different
modes of the SCORPIO multipurpose
spectrograph~\citep{afa05:Afanasieva_n}, at which the CCD systems with
EEV CCD42-40 and CCD42-90 detectors, are used as registration
systems. For instance, in the direct image mode, a flux from the
moonless sky background greater than $100 e^-$ is achieved in the
green spectral region (the {\it V} filter) in 10-second exposure and
in the red one (the {\it R} filter)---in 5-second exposure; and as it
follows from Fig.~3, minimum distortions of the
counts statistics and maximum signal-to-noise ratio are achieved in
the flux range of $10^2$--$10^3 e^-$. As can be seen from
Fig.~4, with exposures greater than 100~s, the
{\it S/N} ratio does not almost increase. In order to extend the
detection threshold, the most correct observational approach is not
the extension of an exposure time but increasing a number of
short-term exposures. After independent processing of each image,
they can be combined and the signal-to-noise ratio will~$\sqrt{N_{\rm
exp}}$ times grow. It is necessary that flat fields of each exposure
do not coincide, which corresponds to image acquisition with an
object shift or observations in different nights. Otherwise, an
inhomogeneity value ceases to be random and its contribution does not
decrease after combining images. It should be noted that such a way
of observations not only increases the {\it S/N} but also allows us
to efficiently remove cosmic rays in the combined image if the number
of exposures is more than three. In this case, a median and robust
average are calculated in each channel. Such an algorithm helps to
eliminate cosmic rays without distorting the spectrum of spatial
image frequencies.

\section{CONCLUSIONS}
For the first time a statistics of counts at the output of a CCD
system has been studied using measurements of a Fano factor.
Distortions of statistics when acquiring faint objects has also been
investigated. It is shown that:
\begin{list}{}{
\setlength\leftmargin{2mm} \setlength\topsep{2mm}
\setlength\parsep{0mm} \setlength\itemsep{2mm} } \item(1) For the
existing CCD readout channel designed in the Advanced Design
Laboratory, the deviation from the Poisson statistics is observed for
fluxes weaker than $100 e^-$ with a readout noise of $3 e^-$. The CCD
system can be regarded as almost ``ideal'' in the flux range of
$10^2$--$10^4 e^-$. \item(2) Distortions of counts statistics at
various spatial frequencies are small---the Fano factor increases
1.5~times at scales of 10 to 100 pixels after the removal of fringes.
\item(3) The Fano factor for the counts of a faint object with a
contrast of about $1$ depends on the sky background level, and the
increase of exposure time does not cause the increase of a
signal-to-noise ratio due to errors in the flat field determination.
\end{list}
The results derived can be used to choose optimum exposure ranges in
detecting extremely low signals when one is registering background
radiation and spectra of distant faint galaxies.

\begin{acknowledgements}
The author is thankful to Dr.~Sci.~(Phys.-Math.),
Prof.~V.~L.~Afanasiev for the goal setting and valuable discussions
and to the staff of the Advanced Design Laboratory of SAO RAS for
their help in performing this investigation. The author thank the
reviewer for critical comments. The data from the ASPID database 
were used in the paper.
Observations at the 6-m SAO RAS telescope are carried out with the
financial support of the Ministry of Education and Science of the
Russian Federation (agreement No.~14.619.21.0004, project ID
RFMEFI61914X0004).
\end{acknowledgements}

\end{document}